\documentclass{llncs}
\usepackage{makeidx}  
\usepackage{amsfonts,amssymb}

\begin{document}

\title{Improved Approximability Result for Test Set with Small Redundancy}
\titlerunning{Improved Approximability Result}  
%
\author{Peng Cui}
\authorrunning{Peng Cui}   
\institute{Renmin University of China, Beijing 100872, China\\
\email{cuipeng@ruc.edu.cn}}

\maketitle              

\begin{abstract}
Test set with redundancy is one of the focuses in recent
bioinformatics research. Set cover greedy algorithm (SGA for short)
is a commonly used algorithm for test set with redundancy. This
paper proves that the approximation ratio of SGA can be
$(2-\frac{1}{2r})\ln n+\frac{3}{2}\ln r+O(\ln\ln n)$ by using the
potential function technique. This result is better than the
approximation ratio $2\ln n$ which directly derives from set
multicover, when $r=o(\frac{\ln n}{\ln\ln n})$, and is an extension
of the approximability results for plain test set.
\end{abstract}
\section{Preliminaries}

\subsection{Test Set Problems}
Test set problems arise in pattern recognition, machine learning,
and bioinformatics. Test set is NP-hard. The algorithms used in
practice include simple "greedy" algorithms, branch and bound, and
Lagrangian relaxation. The "greedy" algorithms can be implemented by
set cover criterion or by information criterion, and the average
performances of the two types of "greedy" algorithms are virtually
the same in practice\cite{ms}. Test set is not approximable within
$(1-\varepsilon)\ln n$ for any $\varepsilon>0$ unless $NP\subseteq
DTIME(n^{\log\log n})$\cite{bh,bdk}.

Recently, the precise worst case analysis of the two type "greedy"
algorithms has been accomplished. The authors of \cite{bdk} designed
a new information type algorithm, information content heuristic (ICH
for short), and proved its approximation ratio $\ln n+1$, which
almost matches the inapproximability result. The author of \cite{c}
proved that the approximation ratio of set cover greedy algorithm
(SGA for short) can be $1.14\ln n$, and showed a lower bound
$1.0007\ln n$ of the approximation ratio of this algorithm.

Test set with redundancy, which can be regarded as a special case of
set multicover\footnote{This paper considers the case each subset
can be selected only once, which is called constrained set
multicover in: Vazirani V V. Approximation Algorithms. Springer,
2001. 108-118.}, captures the requirement of redundant
distinguishability in the string barcoding problem\cite{dk} and the
minimum cost probe set problem\cite{bc} in bioinformatics.

The input of test set with redundancy $r\in Z^+$ consists of a set
of items $S$ with $|S|=n$, a collection of subsets (called tests) of
$S$, $\mathcal T$. An item pair is a set of two different items. A
test $T$ differentiates item pair $a$ if $|T\cap a|=1$. A family of
tests $\mathcal T'\subseteq \mathcal T$ is a $r$-test set of $S$ if
each item pair is differentiated by at least {\it different} $r$
tests in $\mathcal T'$. The objective is to find out the $r$-test
set of minimum cardinality. 1-test set is simply abbreviated to test
set.

\begin{definition} [Test Set with Redundancy $r$]
\footnote{In Definition 1, we suppose there are no two tests $T_{1}$
and $T_{2}$ satisfying $T_{1}=S-T_{2}$.}\\
\indent{\bf Input:} $S$, $\mathcal T$;\\
\indent{\bf Feasible Solution:} $r$-test set $\mathcal T'$,
$\mathcal T'\subseteq\mathcal T$;\\
\indent{\bf Measure:} $|\mathcal T'|$;\\
\indent{\bf Goal:} minimize.
\end{definition}

We use $a\perp T$ to indicate the fact that $T$ differentiates $a$
and use $\perp (a,\mathcal T)$ to represent the number of tests in
$\mathcal T$ that differentiate $a$. We give the following two facts
without proof. If $\mathcal T$ is a $r$-test set, then $|\mathcal
T|\ge \log_{2} n$. If $\mathcal T$ is a minimal $r$-test set, then
$|\mathcal T|\le r(n-1)$.

\subsection{Set Cover Greedy Algorithm}
Test set with redundancy can be reduced to set multicover in a
natural way. Let $(S,\mathcal T)$ be an instance of test set with
redundancy $r$, we construct an instance $(U,\mathcal C)$ of set
multicover with coverage requirement $r$, with $U=\{\{i,j\}|i,j\in
S,i\ne j\}$, and
$$\mathcal C=\{c(T)|T\in \mathcal T\},c(T)=\{\{i,j\}|i\in T,j\in S-T\}$$

Clearly,  $\mathcal T'$ is a $r$-test set iff $\mathcal
C'=\{c(T)|T\in \mathcal T'\}$ is a $r$-set cover of $U$.

SGA runs the same way as the greedy algorithm for set multicover. We
say an item pair $a$ is {\it alive} if it is differentiated by fewer
picked tests than $r$. In each iteration, the algorithm picks, from
the currently unpicked tests, the tests differentiated most
undifferentiated alive item pairs. Formally, SGA can be described as:\\

\noindent{\bf Algorithm.} SGA\\
{\bf Input:} $S$,$\mathcal T$;\\
{\bf Output:} a $r$-test set of $S$;\\
{\bf begin}\\
\indent$\bar\mathcal T\leftarrow\varnothing$;\\
\indent{\bf while} $\#(\bar\mathcal T)>0$ {\bf do}\\
\indent\indent select $T$ in $\mathcal T-\bar\mathcal T$
minimizing $\#(\bar\mathcal T\cup\{T\})$;\\
\indent\indent $\bar\mathcal T\leftarrow\bar\mathcal T\cup\{T\}$;\\
\indent{\bf endwhile}\\
\indent{\bf return}$\bar\mathcal T$.\\
{\bf end}

\begin{definition} [Partial $r$-Test Set and Differentiation Measure]\\
\indent We call $\bar\mathcal T$ the partial $r$-test set. The
differentiation measure of $\bar\mathcal T$ is defined as
$\#(\bar\mathcal T)=\sum_{a}\max (r-\perp (a, \bar\mathcal T),0),$
and the differentiation measure of $T$ related to $\bar\mathcal T$
is defined as $\#(T,\bar\mathcal T)=\#(\bar\mathcal
T)-\#(\bar\mathcal T\cup\{T\})$. Denote
$\#_{0}=\#(\varnothing)=rn(n-1)/2$.
\end{definition}

The greedy algorithm for set multicover has approximation ratio $H_{
N}$, where $N=|U|$\cite{rv}. Using the natural reduction, we
immediately obtain the approximation ratio $2\ln n$ of SGA. Using
the standard multiplicative weights argument, we can obtain another
approximation ratio $\ln \#_{0}-\ln m^*+1$ of SGA, where $m^*$ is
the size of the optimal $r$-test set(See Lemma 19 in \cite{gw}).

The authors of \cite{bds} designed a randomized multi-step rounding
algorithm (RND for short) for set multicover, and the expectation of
the approximation ratio is approximately no more than $\ln N-\ln r$.
Experiments on test set show when $r$ is small, SGA performs better
than RND, and when $r$ is near to or more than $n$, RND performs
better than SGA\cite{dk}.

\subsection{Our Method and Result}
In \cite{y}, Young addresses "oblivious rounding" technique to get
another proof the of the well-known approximation ratio $\ln n+1$ of
the greedy algorithm for set cover. He observes the number of
elements uncovered is an "potential function" and the approximation
algorithm only needs to drive down the potential function at each
step.

Arora et al. present a simple meta algorithm that unifies many
disparate algorithms and drive them as instantiations of the meta
algorithm\cite{ahk}. They call the meta algorithm multiplicative
weights method, and suggest it is viewed as a basic tool for
designing algorithms.

This paper proves that the approximation ratio of SGA can be
$(2-\frac{1}{2r})\ln n+\frac{3}{2}\ln r+O(\ln\ln n)$ by applying the
potential function technique. This result is better than the
approximation ratio $2\ln n$ which directly derives from set
multicover, when $r=o(\frac{\ln n}{\ln\ln n})$, and is an extension
of the approximability results for plain test set. The analysis of
this algorithm fits in the framework of multiplicative weights
method.

In Section 2, the authors analyze the phenomenon of "differentiation
repetition" of test set with redundancy and apply the potential
function technique to prove improved approximation ratio of SGA.
Section 3 is some discussions.

\section{Proof of Our Result}

\subsection{Differentiation Repetition}
Practitioners of test set problems are aware of the phenomenon that
the number of times for which the item pairs are differentiated
tends to be more than the requirement. In another word, item pairs
differentiated for small number of times are quite "sparse",
especially when $m^*$ is small.

The author of \cite{c} investigates this unique characteristic of
test set quantitatively. He analyzes the distribution of times for
which item pairs are differentiated, especially the relationship
between the differentiation distribution and the size of the optimal
test set. The following lemma on test set with redundancy can be
obtained as a corollary.

\begin{lemma}
Let $\mathcal T^*$ be an optimal $r$-test set, and $m^*=|\mathcal
T^*|$, then at most $2n\log_{2}nm^{*{r-1}}$ item pairs are
differentiated by exactly $r$ test in $\mathcal T^*$.
\end{lemma}

\subsection{Improved Approximation Ratio}
In this subsection, the authors apply the potential function
technique to prove improved approximation ratio of SGA. We note the
decrease of the potential function can be "accelerated" in the
beginning phase of SGA. Our proof is based on the technique to
balance the potential function by appending a negative term to the
differentiation measure.

\begin{lemma}
Given an instance $(S,\mathcal T)$ of test set with redundancy $r$,
let $\mathcal T^*$ be an optimal $r$-test set, $m^*=|\mathcal T^*|$,
and $\#_{B}$ is the number of item pairs differentiated by exactly
$r$ tests in $\mathcal T^*$, then the size of the solution returned
by SGA is no more than $(\ln
\#_{0}-\frac{1}{r+1}\ln{\frac{\#_{0}}{\#_{B}}}+\frac{r}{r+1}\ln
(r+1)+1)m^*+1$.
\end{lemma}

\begin{proof}
Clearly, there is a partial $r$-test set $\mathcal T_{1}$ such that
$\#(\mathcal T_{1})\ge\#_{B}$, but after selecting the next test
$\tilde T$, $\#(\mathcal T_{1}\cup\{\tilde T\})<\#_{B}$. Let the set
of selected tests after selecting $\tilde T$ until the algorithm
stops is $\mathcal T_{2}$. Then the returned $r$-test set is
$\mathcal T'=\mathcal T_{1}\cup\{\tilde T\}\cup \mathcal T_{2}$. Let
$k=\frac{r}{r+1}\ln \frac{(r+1)\#_{0}}{\#_{B}}m^*$.

Define the potential function as
$$f(\bar\mathcal T)=(\#(\bar\mathcal T)-\frac{r}{r+1}
\#_{B})(1-\frac{{r+1}}{r} \frac{1}{m^*})^{k-|\bar\mathcal T|}.$$

Then
$$f(\varnothing)=(\#_{0}-\frac{r}{r+1}\#_{B})(1-\frac{{r+1}}{r}\frac{1}{m^*})^{k}
<\#_{0}/(\frac{(r+1)\#_{0}}{\#_{B}})=\frac{\#_{B}}{r+1}.
$$

Given $\bar\mathcal T$, let $\bar p$ denote the probability
distribution on tests in $\mathcal T^*-\bar\mathcal T$: draw one
test uniformly from $\mathcal T^*-\bar\mathcal T$. For any $T\in
\mathcal T^*-\bar\mathcal T$, the probability of drawing $T$ is
$\bar p(T)=\frac{1}{|\mathcal T^*-\bar\mathcal T|}$.

For any item pair $a$,
$$\sum_{T\in \mathcal T^*-\bar\mathcal T:a\perp T}{\bar p(T)}
=\frac{\perp (a,\mathcal T^*-\bar\mathcal T)}{|\mathcal
T^*-\bar\mathcal T|} \ge\frac{\perp (a,\mathcal T^*)-\perp
(a,\bar\mathcal T)}{m^*}.$$

Since $\perp (a,\mathcal T^*)\ge r$,
$$\sum_{T\in \mathcal T^*-\bar\mathcal T:a\perp T}{\bar p(T)}\ge
\frac{r-\perp(a,\bar\mathcal T)}{m^*}.$$

If $\perp(a,\mathcal T^*)\ge r+1$,
$$\sum_{T\in \mathcal
T^*-\bar\mathcal T:a\perp T}{\bar p(T)}\ge
\frac{r-\perp(a,\bar\mathcal T)+1}{m^*}.$$

By the definition of $f(\bar\mathcal T)$ and the facts $\bar p(T)\ge
0$ and $\sum_{T\in \mathcal T^*}{\bar p(T)}=1$,
\begin{eqnarray*}
&&\min_{T\in \mathcal T-\bar\mathcal T}{f(\bar\mathcal T\cup \{T\})}\\
&&\le\min_{T\in \mathcal T^*-\bar\mathcal T}{f(\bar\mathcal T\cup \{T\})}\\
&&\le\sum_{T\in \mathcal T^*-\bar\mathcal T}{(\bar
p(T)}f(\bar\mathcal T\cup\{T\}))\\
&&=(\#(\bar\mathcal T)-\frac{r}{r+1}\#_{B}-\sum_{T\in \mathcal
T^*-\bar\mathcal T}({\bar p(T)\#(T,\bar\mathcal
T)}))(1-\frac{{r+1}}{r} \frac{1}{m^*})^{k-|\bar\mathcal T|-1}\\
&&=(\#(\bar\mathcal
T)-\frac{r}{r+1}\#_{B}-\sum_{alive\,a}{\sum_{T\in \mathcal
T^*-\bar\mathcal T:a\perp T}{\bar p(T)}})(1-\frac{{r+1}}{r}
\frac{1}{m^*})^{k-|\bar\mathcal T|-1}
\end{eqnarray*}
and
\begin{eqnarray*}
&&\sum_{alive\,a}{\sum_{T\in
\mathcal T^*-\bar\mathcal T:a\perp T}{\bar p(T)}}\\
&&\ge\sum_{a:\perp(a,\mathcal T^*)=r}{\frac{r-\perp(a,\bar\mathcal
T)}{m^*}}+\sum_{a:\perp(a,\mathcal T^*)\ge
r+1}{\frac{r-\perp(a,\bar\mathcal T)+1}{m^*}}\\
&&=\sum_{alive\,a}{\frac{r-\perp(a,\bar\mathcal T)+1}{m^*}}
-\sum_{a:\perp(a,\mathcal T^*)=r}{\frac{1}{m^*}}\\
&&=\frac{1}{m^*}\sum_{alive\,a}{((r-\perp(a,\bar\mathcal
T))\frac{r-\perp(a,\bar\mathcal T)+1}{r-\perp(a,\bar\mathcal
T)})}-\frac{1}{m^*}\#_{B}\\
&&\ge\frac{{r+1}}{r}\frac{1}{m^*}(\#(\bar\mathcal
T)-\frac{r}{r+1}\#_{B}).
\end{eqnarray*}

Hence
$$\min_{T\in \mathcal T-\bar\mathcal T}{f(\bar\mathcal T\cup
\{T\})}\le(\#(\bar\mathcal
T)-\frac{r}{r+1}\#_{B})(1-\frac{{r+1}}{r}\frac{1}{m^*})(1-\frac{{r+1}}{r}
\frac{1}{m^*})^{k-|\bar\mathcal T|-1}=f(\bar\mathcal T).$$

For partial $r$-test set $\bar\mathcal T$, the algorithm selects $T$
in $\mathcal T-\bar\mathcal T$ to minimize $f(\bar\mathcal
T\cup\{T\}$). Therefore, $f(\mathcal T_{1})\le f(\varnothing)\le
\frac{\#_{B}}{r+1}$.

By definition of $\mathcal T_{1}$,
$$f(\mathcal T_{1})\ge
(\#_{B}-\frac{r}{r+1} \#_{B})(1-\frac{{r+1}}{r}
\frac{1}{m^*})^{k-|\mathcal
T_{1}|}=\frac{\#_{B}}{r+1}(1-\frac{{r+1}}{r}
\frac{1}{m^*})^{k-|\mathcal T_{1}|}.$$

Therefore, $(1-\frac{{r+1}}{r} \frac{1}{m^*})^{k-|\mathcal
T_{1}|}<1$, and $|\mathcal T_{1}|<k$.

We can easily prove $|\mathcal T_{2}|<(\ln \#_{B}+1)m^*$ by natural
reduction to set multicover. When the algorithm stops, the size of
the returned solution is
$$|\mathcal T'|=|\mathcal T_{1}|+|\mathcal T_{2}|+1<(\ln
\#_{0}-\frac{1}{r+1}\ln{\frac{\#_{0}}{\#_{B}}}+\frac{r}{r+1}\ln
(r+1)+1)m^*+1.$$ \qed
\end{proof}

\begin{theorem}
The approximation ratio of SGA for test set with redundancy $r$ can
be $(2-\frac{1}{2r})\ln n+\frac{3}{2}\ln r+O(\ln\ln n)$.
\end{theorem}
\begin{proof}
Let $\rho_1=\ln \#_{0}-\ln m^*+1$, and $\rho_2=\ln
\#_{0}-\frac{1}{r+1}\ln{\frac{\#_{0}}{2n\log_{2}nm^{*{r-1}}}}+\frac{r}{r+1}\ln
(r+1)+1$. Then $\rho_1$ is an upper bound of the approximation ratio
(\cite{gw}), and $\rho_1$ is also an upper bound of the
approximation ratio by Lemma 1 and Lemma 2.

For fixed $r$ and $n$, $\rho_1$ is a decreasing function of $m^*$,
and $\rho_2$ is an increasing function of $m^*$.
$min(\rho_1,\rho_2)$ is maximized when $\rho_1=\rho_2$. This leads
to $\ln m^*=\frac{1}{2r}\ln n-\frac{1}{2}\ln r-O(\ln\ln n)$, which
implies $min(\rho_1,\rho_2)\le(2-\frac{1}{2r})\ln n+\frac{3}{2}\ln
r+O(\ln\ln n)$.\qed
\end{proof}

\section{Discussions}
In this paper, the authors show new approximability result for test
set with small redundancy, which is better than approximation ratio
which directly derives from set multicover. It seems that ICH can
not be generalized to test set with redundancy $r>1$. This situation
raises an interesting problem if the approximation ratio of test set
with redundancy can be pushed to the matching bound $\ln n+1$ of
plain test set.


\end{document}